\documentclass[secnumarabic,amssymb, nobibnotes, aps, prl,reprint]{revtex4-1}
\usepackage{graphicx}	
\usepackage{siunitx}
\usepackage{amsmath}
\usepackage[caption = false]{subfig}
\usepackage{epstopdf}
\begin{document}
\title{Bose-Einstein Condensation of Long-Lifetime Polaritons in Thermal Equilibrium}
\author{Yongbao Sun,$^{1}{}^\ast$  Patrick Wen,$^{1}$ Yoseob Yoon,$^{1}$ Gangqiang Liu,$^{2}$ Mark Steger,$^{2}$\\ Loren N. Pfeiffer,$^{3}$ Ken West,$^{3}$ David W. Snoke,$^{2}{}^\ast$ and Keith A. Nelson,$^{1}$}
\affiliation{${}^1$Department of Chemistry and Center for Excitonics, Massachusetts Institute of Technology, 77 Massachusetts Avenue, Cambridge, MA 02139, USA \\ 
${}^2$Department of Physics, University of Pittsburgh, 3941 O'Hara St., Pittsburgh, PA 15218, USA\\
${}^3$Department of Electrical Engineering, Princeton University, Princeton, NJ 08544, USA
}

\begin{abstract} 
The experimental realization of Bose-Einstein condensation (BEC) with atoms and quasiparticles has triggered wide exploration of macroscopic quantum effects. Microcavity polaritons are of particular interest because quantum phenomena such as BEC and superfluidity can be observed at elevated temperatures. However, polariton lifetimes are typically too short to permit thermal equilibration. This has led to debate about whether polariton condensation is intrinsically a nonequilibrium effect. Here we report the first unambiguous observation of BEC of optically trapped polaritons in thermal equilibrium in a high-Q microcavity, evidenced by equilibrium Bose-Einstein distributions over broad ranges of polariton densities and bath temperatures. With thermal equilibrium established, we verify that polariton condensation is a phase transition with a well defined density-temperature phase diagram. The measured phase boundary agrees well with the predictions of basic quantum gas theory. 
\end{abstract}
\maketitle

The realization of exciton-polariton condensation in semiconductor microcavities from liquid helium temperature \cite{Kasprzak2006, Balili2007} all the way up to room temperature \cite{Christopoulos2007, Kena-Cohen2010, Plumhof2014}  presents great opportunities both for fundamental studies of many-body physics and for all-optical devices on the technology side. Polaritons in a semiconductor microcavity are admixtures of the confined light modes of the cavity and excitonic transitions, typically those of excitons in semiconductor quantum wells (QWs) placed at the antinodes of the cavity. Quantum effects such as condensation \cite{Kasprzak2006, Balili2007, Christopoulos2007, Kena-Cohen2010, Plumhof2014}, superfluidity \cite{Amo2009}, and quantized vortices \cite{Lagoudakis2008, Lagoudakis2009, Sanvitto2010, Nardin2011, Tosi2012} have been reported. The dual light-matter nature permits flexible control of polaritons and their condensates, facilitating applications in quantum simulation. It is also straightforward to measure the spectral functions, $A(k,\omega)$, of polaritons, which can provide insights into the dynamics of many-body interactions in polariton systems. For cold atoms, the equilibrium occupation numbers can be measured \cite{Ensher1996}, but the spectral function is not readily accessible. Observations of non-Hermitian physics \cite{Gao2015} and phase frustration \cite{Baboux2015} have shown that polaritons are an important complement to atomic condensates. 

However, in most previous experiments, the lifetime of the polaritons in microcavities has been 30 ps or less \cite{Wertz2010} due to leakage of the microcavity. Thus, although there have been claims to partial thermalization of polaritons \cite{Deng2006, Kasprzak2008}, no previous work has unambiguously shown a condensation in thermal equilibrium, leading to the common description of polariton condensates as ``nonequilibrium condensates'' \cite{ Byrnes2014, Dominici2015, Sanvitto2016}. The theory of nonequilibrium condensation is still an active field \cite{Szymanska2006,Wouters2010, Keeling2011, Smith2012}.  Although polariton experiments and theory have shown that a great number of canonical features of condensation persist in nonequilibrium, e.g., superfluid behavior \cite{Wouters2010, Keeling2011}, some aspects may not \cite{Janot2013, Altman2015},  and debates persist over whether polariton condensates can be called Bose-Einstein condensates \cite{Butov2007, Butov2012, Deveaud-Pledran2012}, in part related to the question of whether polariton condensation is intrinsically a nonequilibrium effect. 
It is thus of fundamental importance to investigate whether polariton condensates can reach thermal equilibrium.  Of course, strictly speaking, BEC cannot occur in an ideal infinite 2D system, but it has been shown \cite{Holzmann2007, Berman2008} that a 2D Bose gas in a large but finite trap has the same threshold behavior as a 3D Bose gas in a finite trap of the same type. We can thus talk of an equilibrium BEC in 2D and 3D finite trapped systems using the same language.  

\vspace*{0.35cm}
\noindent\textbf{Trapping polaritons in a high-$Q$ microcavity} 

\noindent The main challenge in reaching full thermalization in polariton systems is to achieve a very long polariton lifetime, longer than their thermalization time. The thermalization time for the polariton gas was estimated to be at least 40 ps for polaritons that are mostly exciton-like \cite{Deng2006}, and can be even longer for more photon-like polaritons which are less interactive. However, most samples used in previous experiments have polariton lifetimes on the order of a few picoseconds. This suggests that an improvement of the cavity $Q$ by at least an order of magnitude is needed, which is not trivial for GaAs fabrication technology. We have succeeded at this by growing a GaAs-based high-Q microcavity structure by molecular beam epitaxy. The main change from samples used in previous experiments \cite{Balili2007} was to double the number of the quarter-wavelength layers in the distributed Bragg reflectors (DBRs) that make up the mirrors of the cavity; the detailed recipe and the difficulties involved in fabricating a long lifetime sample are described in the Supplementary Information. The new microcavity structure has a $Q$ of $\sim$320,000 and a cavity photon lifetime of $\sim$135 ps. This corresponds to a polariton lifetime of 270 ps at resonance, which has been confirmed by the long-range (millimeter scale) propagation of polaritons created through either resonant \cite{Steger2015} or non-resonant \cite{Steger2013} optical excitation. 

Due to the light effective mass and inefficient scattering with phonons, polaritons can propagate over long distances, up to millimeters when they are mostly photonic \cite{Steger2013,Steger2015}. In order to guide them toward equilibrium with a specified location and geometry, we made a spatial trap.  We created an annular optical trap to localize polaritons under non-resonant excitation. This method has been used previously in several experiments to confine polaritons \cite{Tosi2012_2,Cristofolini2013, interactions, Askitopoulos2013, Askitopoulos2015}. The excitation pattern on the microcavity is an annulus with a diameter of 38 $\mu$m, as shown in Fig.~1a. In Fig.~1b, we show the normalized light intensity plot for the $x=0$ slice of the ring pattern in Fig.~1a. The light intensity in the center is nearly negligible; as discussed above, it gives rise to a nearly flat potential for the polaritons in the center of the ring, in which the variation in energy is much less than $k_BT$.
\begin{figure}[htbp]
 \centering
   \includegraphics[width=0.5\textwidth]{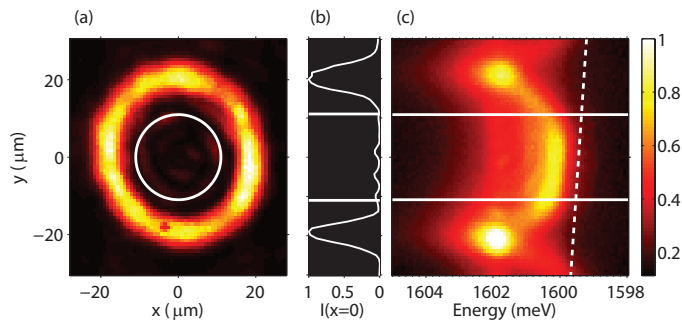}
    \caption{ (color online). (a) Reflection of the excitation beam from the sample surface. The white circle indicates the region of the sample that is observed in photoluminescence (PL) imaging measurements after spatial filtering. (b) Normalized excitation light intensity along the $x=0$ line through the center of the excitation ring pattern shown in (a). (c) Spectrally resolved PL along $x=0$. The PL within the solid white lines is collected and imaged onto the spectrometer CCD in the far-field geometry for the polariton distribution measurements. The dashed white line indicates the photon energy gradient deduced from the low-density spectrum.}
 \end{figure}
We also plotted the intensity of the photoluminescence (PL) emitted by lower polaritons as a function of both the PL energy and the sample position for the case of moderate pump power below the condensation threshold in Fig.~1c. The white dashed line indicates the emission energies at very low pump powers; the slope of this line arises from the wedge of the cavity thickness, which causes a gradient in the cavity photon energy. At the pump region, there is a blue shift of the polariton energy due to their interactions with each other as well as from repulsive interactions between polaritons and excitons and free carriers. As seen in Fig.~1c, the barrier is not constant around the ring, varying by about 1 meV from one side to the other due to inhomogeneity in the pump intensity. The barrier is slightly wider than the laser profile, because excitons propagate up to 10 $\mu$m. The potential landscape is nearly flat in the region from $-$11 $\mu$m to 11 $\mu$m indicated by the white circle in Fig.~1a and the horizontal lines in Fig.~1b and 1c.  PL was collected from only this region for determination of the polariton energy distribution as discussed below. The nearly flat potential profile corresponds to a constant density of states in 2D. Additionally, a nearly homogeneous distribution was established in the field of view, as evidenced by little change in the energy-resolved emission intensities in Fig.~1c (see Fig. S6 in Ref.~\cite{interactions} for a direct measurement of spatial profiles under similar conditions). 

Polaritons are generated in the pump region and stream away in all directions.  However, only polaritons that propagate into the center of the trap can meet and interact,  leading to the accumulation of the densities high enough for condensation. A near-field image of the sample was projected onto a spatial filter at a reconstructed real-space plane of the sample surface to select only the PL from inside the trap (within the white circle in Fig.~1a), and a far-field image of the PL that passed through the spatial filter was projected onto an imaging spectrometer, giving the intensity of the PL, $I$, as a function of both the in-plane wavevector component, $k_{||}$, and the corresponding energy, $E(k_{||})$. The dispersion $E(k_{||})$ is given in the supplementary information. Finally, $I[E(k_{||})]$ was converted into the number of polaritons, $N[E(k_{||})]$ (hereafter simply $N(E)$),  by the use of one single efficiency factor throughout the experiments. Detailed information about how to determine the efficiency factor can be found in Ref.~\cite{interactions}. Crucially, the same efficiency factor was used for all the distributions so that the absolute occupation numbers of different distributions could be compared.

\vspace*{0.35cm}
\noindent\textbf{Varying the polariton gas from nonequilibrium to equilibrium} 

\noindent To see the effect of interactions on thermalization, $N(E)$ was measured at two different cavity detunings, $\delta=-5$ meV and $\delta=0$ meV, for a series of pump powers. The detuning $\delta$ is the energy difference between the cavity resonance and exciton energy at $k_{||}=0$.  Changing the detuning changes the underlying excitonic fraction of the polaritons, which governs the strength of their interactions. Positive values of detunings indicate polaritons are more exciton-like, while negative values of detunings give polaritons which are mostly photon-like.  Here $\delta = 0$ meV and $\delta = -5$ meV correspond to excitonic fractions of 50\% and 30\%, respectively.  This indicates that the polaritons with $\delta=-5$ meV have interactions which are weaker by a factor of 3 than those at $\delta = 0$ meV, and less well thermalization is expected. 

 \begin{figure}[htbp]
 \centering
   \includegraphics[width=0.42\textwidth]{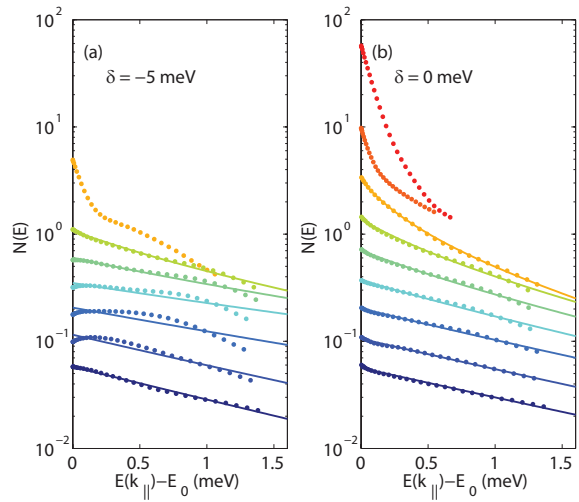}
    \caption{ (color online).  Energy distributions of polaritons in the center of the trap at (a) $\delta = -5$ meV (b) $\delta = 0$ meV at a bath temperature of $T_{bath} = 12.5$ K at different pump powers (see supplementary information for values). The solid curves are best fits to the equilibrium Bose-Einstein distribution in Eq.~(\ref{eq:BEdistrib}). The fitted values of $T$ and $\mu$ are shown in Fig.~3. The power values from low to high are 0.12, 0.24, 0.45, 0.71, 0.93, 1.07, 1.10, 1.12 and 1.14 times of the threshold values $P_{BE}$, which are 382 mW and 443 mW for detunings $\delta$ = −5 meV and $\delta$ = 0 meV, respectively.}
\end{figure}

The measured distributions $N(E)$ at both detunings and various pump powers are shown in Fig.~2. The pump powers are reported in terms of the threshold power, $P_{BE}$, defined below. The sample was immersed in a helium bath that was kept at a temperature $T_{bath} = 12.5$ K for both detuning positions. The measured values of $N(E)$ were fit to a Bose-Einstein distribution, given by
\begin{equation}
\label{eq:BEdistrib}
N_{BE}(E)=\frac{1}{e^{(E-\mu)/k_BT}-1},
\end{equation}
where $T$ and $\mu$ are the temperature and chemical potential of the polaritons, respectively, and $k_B$ is the Boltzmann constant. The ground state ($k_{||}=0)$ of the lower polariton shifts to higher energy as the density increases, due to the repulsive interpolariton interactions. We defined the ground state energy in each case as $E=0$ so that $\mu= 0$ corresponds to the condition for Bose-Einstein condensation. The best fits of the data to $N_{BE}(E)$ were determined using $T$ and $\mu$ as free parameters in nonlinear least-squares regressions, and are shown as solid curves in Fig.~2. 

When the polariton states are negatively detuned and have very weak interactions, the fits to the Bose-Einstein distribution are poor. As seen in Fig.~2(a), for the case of $\delta=-5$ meV, at low density the distribution has a reasonable fit to a Maxwell-Boltzmann distribution (which corresponds to a single exponential, i.e., a straight line on a semi-log plot), but as the polariton density is increased, the distribution is no longer fully thermal. The hump at $E = 0.5$ meV is a manifestation of the bottleneck effect, as was also observed in Ref.~\cite{Kasprzak2008}.  As the density is increased further, a peak occurs which is condensate-like, but the rest of the distribution does not fit the Bose-Einstein functional form in Eq.~(\ref{eq:BEdistrib}), indicating that the polaritons are not in thermal equilibrium.  This behavior is similar to that seen in many other experiments with short-lifetime polaritons, e.g.~Refs. \cite{Kasprzak2006, Kasprzak2008}, and is consistent with a nonequilibrium polariton condensate. The nonequilibrium distribution has been reproduced by numerical solution to the quantum Boltzmann equation \cite{Hartwell2010}. Despite the long cavity lifetime, the photon-like polaritons with weak interactions do not reach thermal equilibrium.

In contrast, $N(E)$ at $\delta=0$ is well described by $N_{BE}(E)$ for all pump powers up to $P=1.1P_{BE}$. At pump powers well below $P_{BE}$, $N(E)$ is well described by a single exponential function, i.e., a Maxwell-Boltzmann distribution.  Between $P=0.9P_{BE}$ and $P=1.1P_{BE}$, an upturn in the distribution at $E=0$ meV is observed, indicating that $N(E)$ deviates from Maxwell-Boltzmann statistics and must be described by a Bose-Einstein distribution with the reduced chemical potential $|\mu/k_BT| < 1$. The fit values of $T$ and $\mu$ used in Fig.~2b are shown in Fig.~3 as the blue symbols. As seen in this figure, when the density is increased, $T$ decreases from around 20 K to a lowest value of $13.9 \pm 0.2$ K and $\mu/k_BT$ smoothly goes from $-2.93 \pm 0.16$ to $-0.28 \pm 0.01$. For pump powers greater than $1.1P_{BE}$, a condensate in the ground state appears. In the weakly-interacting limit, the condensate peak should be delta-function like, which is broadened in the presence of finite-size fluctuations \cite{Eastham2006}.The high-energy tail of the top two curves has the same absolute value, indicating that the population in the excited states saturates when there is a condensate, consistent with a Bose-Einstein condensation phase transition for bosons in thermal equilibrium. 

 \begin{figure}[htbp]
 \centering
   \includegraphics[width=0.5\textwidth]{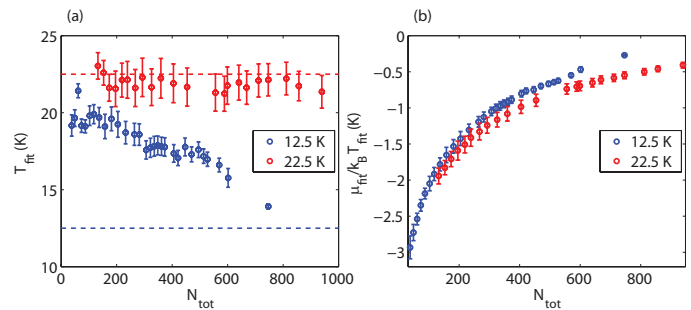}
    \caption{ (color online). (a) Effective temperatures of polaritons for bath temperatures $T=12.5$  K (blue points) and $T=22.5$ K (red points) at different pump powers, extracted by fitting the energy distributions (shown in Fig.~2b for the 12.5 K case) to the equilibrium Bose-Einstein model. The dashed lines indicate the helium bath temperatures. (b) Reduced chemical potential $\alpha=\mu/k_BT$ for bath temperatures $T=12.5$ K (blue points) and $T=22.5$ K (red points) at different pump powers  .}
 \end{figure}

The upturn in the shape of $N(E)$ in the low energy states unambiguously distinguishes $N(E)$ as a Bose-Einstein distribution rather than a Maxwell-Boltzmann distribution. Previous reports using short-lifetime samples \cite{Deng2006,Kasprzak2008} showed fits of $N(E)$ but did not show this behavior; although a condensate peak appeared in some cases, there was not a clear density-dependent evolution from a thermal Maxwell-Boltzmann distribution to a degenerate Bose-Einstein, condensed distribution. Furthermore, short-lifetime samples thermalized only when the microcavity was positively detuned \cite{Deng2006} and the polariton characteristics were mostly exciton-like so that the motion of the polaritons was severely restricted (see Supplementary Information for a detailed discussion). In contrast, the long lifetime polaritons seen here at zero detuning follow Bose-Einstein statistics throughout the phase transition and propagate to fill the trap in spatial equilibrium. 

We emphasize that not only the curvature of the fits in Fig.~2 but also the absolute vertical scale of the fits is constrained by the value of $\mu$. We do not have a free parameter to change the overall intensity scaling factor for each curve. The data points give the absolute occupation numbers as indicated by the vertical scale in addition to the relative occupation numbers at different pump powers. When the value of $\mu$ in the Bose-Einstein distribution is increased toward zero, this increases the absolute value of $N_{BE}(E)$. Thus, the fits are tightly constrained by the requirement that we fit not only the shape of the distribution but also the relative heights of all the curves with only two parameters, $T$ and $\mu$. This constraint is reflected in the very small relative uncertainties in the fit values of $\mu$ shown in Fig.~3b. 

\vspace*{0.35cm}
\noindent\textbf{Phase diagram of polariton Bose condensation} 

\noindent The bath temperature was also varied in the range of 10.0--25.0 K. Good thermalization has been achieved across this range. In Fig. 3, we plot the fitted values of $T$ and $\mu/k_BT$ for different pump powers. As can be seen, when the bath temperature is low, the fit values of $T$ at low densities are much higher than $T_{bath}$ and at higher densities they settle to temperatures slightly above $T_{bath}$, while for a bath temperature of $T = 22.5$ K, the fitted temperatures stay pinned to the bath temperature, within the uncertainty. The chemical potential increases smoothly toward zero in each case as the density is increased.

Now that we have a well defined temperature ranging over which thermal equilibrium is established, it is straightforward to determine the phase diagram of polariton Bose-Einstein condensation. To determine the phase diagram, i.e., to check the scaling law in Eq.~\eqref{eq:rs}, we want to plot the total number of polaritons $N$ as a function of the fit value of $T$ at the threshold. 
\begin{eqnarray}
r_s \sim n^{-1/2} \sim \lambda_{T} \sim \sqrt{\frac{\hbar^2}{mk_BT}}.
\label{eq:rs}
\end{eqnarray}
 We choose the threshold  as the onset of Bose amplification, i.e., $N(k_{||}=0)=1$.  

Based on this methodology, $N_{BE}$ were determined for a series of $T_{bath}$ values ranging from 10.0 K to 25.0 K. The fit values of $T$ at the onset of Bose-Einstein statistics are plotted in Fig.~4a, showing the general trend of $T_{BE}$ slightly higher than $T_{bath}$, as discussed earlier in the text. The relative deviation is highest at low bath temperature, when the heat capacity of the sample is lowest, allowing the local sample temperature to rise more due to the laser heating. 

The phase diagram of Bose-Einstein transition, i.e., the relation of  $N_{BE}$ to $T_{BE}$ is shown in Fig.~4b. The black line is the best fit of a linear proportionality . Within the uncertainty, the data are consistent with a linear increase  in threshold $T$ with $N_{BE}$, consistent with the expected phase boundary of a weakly interacting boson gas in two dimensions implied by the relation \eqref{eq:rs}. This line can be viewed as a phase boundary: above the line, the gas is quantum-degenerate, and below it, the gas is classical. 
 \begin{figure}[htbp]
 \centering
   \includegraphics[width=0.5\textwidth]{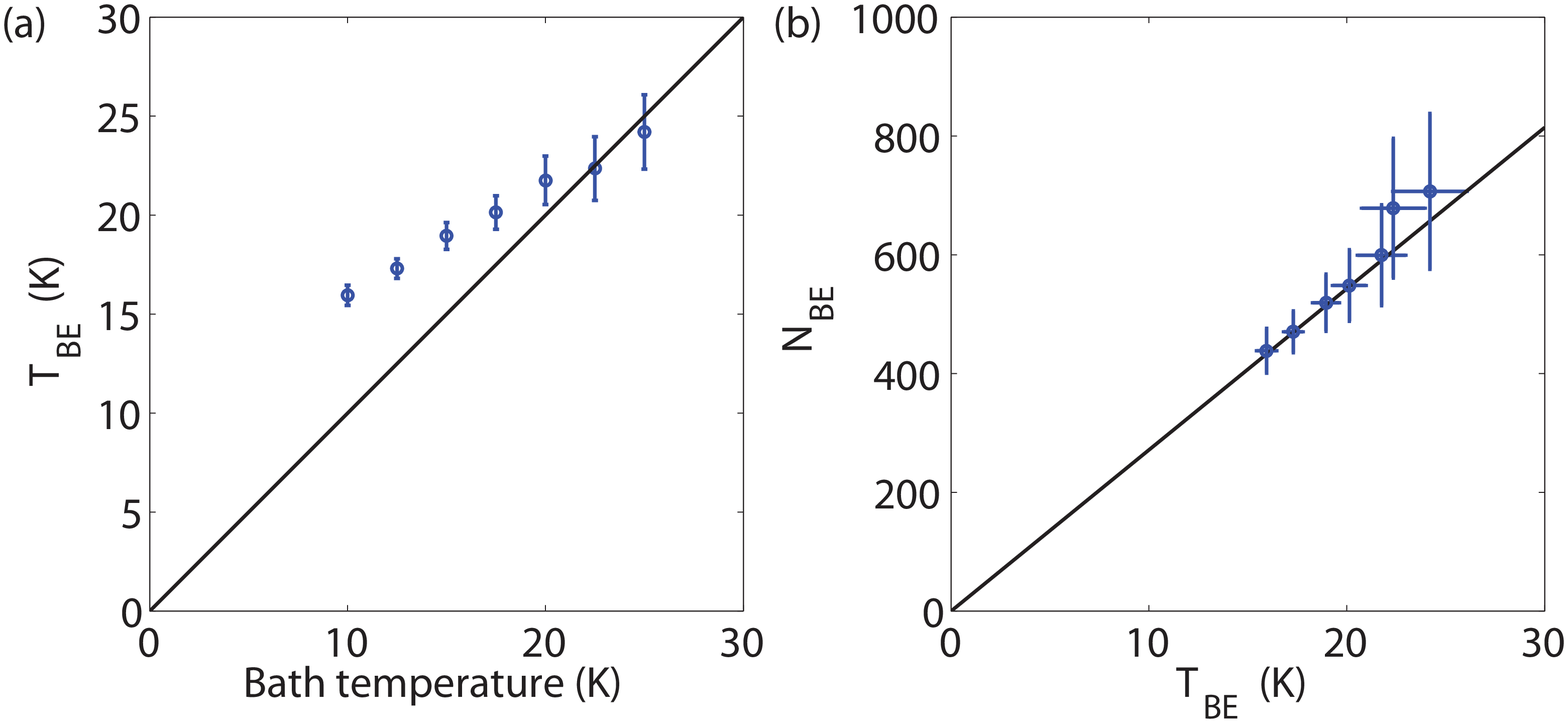}
    \caption{ (color online). (a) The critical temperature as a function of lattice temperature for $\delta = 0$ meV. The solid black line indicates $T_{BE}=T_{\textrm{bath}}$. (b) \textit{Phase diagram of the transition to a degenerate Bose gas}. The solid black line shows the best fit of a linear relation $N_{BE} \propto  T_{BE}$.}
 \end{figure}

It has been a longstanding assumption that the Bose condensation effects seen in polariton systems are a direct result of the quantum nature of the system when $r_s \sim \lambda_{T}$, but up to now it has not been possible to directly test this.  By using high-quality microcavities with lifetimes over an order of magnitude longer than those of previous samples, polaritons within a two-dimensional flat optical trap are seen to unambiguously show thermal Bose-Einstein statistics. This clearly distinguishes polariton condensation from the conventional lasing effect in semiconductor materials. 

Now that we have samples in which true equilibrium can be established, more new experiments are possible to test theoretical predictions of interacting Bose gases which have been elusive in cold atom experiments.  Additionally, studies can be conducted of the excitation spectrum of the interacting Bose gas, and of the crossover from 2D to 1D equilibrium which can be controlled by spatial shaping of the excitation light to make tailored potential energy landscapes. Characteristics of the nonequilibrium state can also be studied systematically by varying the cavity detuning to control the polariton interaction strength and excitation profile to tailor the potential landscape. Dynamical relaxation into the equilibrium state can also be studied by using pulsed rather than c.w.~excitation followed by time-resolved measurements, as well as the coherence properties as the system passes through the Berezinsky-Kosterlitz-Thouless transition.  The results are also encouraging for applications in quantum simulation of condensed matter system that exploit equilibrium BEC properties \cite{misc}. 

This work was supported as part of the Center for Excitonics, an Energy Frontier Research Center funded by the US Department of Energy, Office of Science, Office of Basic Energy Sciences under Award Number DE-SC0001088, and by the  National  Science  Foundation under grants PHY-1205762 and DMR-1104383, and by the Gordon and Betty Moore Foundation through the EPiQS initiative Grant GBMF4420, and by the National Science Foundation MRSEC Grant DMR-1420541.

\bibliography{thermalization}

\newpage~\newpage
\setcounter{figure}{0}
\renewcommand{\thefigure}{S\arabic{figure}}
\subsection*{Supplementary Information for ``Bose-Einstein Condensation of Long-Lifetime Polaritons in Thermal Equilibrium''}

\begin{figure}[htbp]
\centering
  \includegraphics[width=0.5\textwidth]{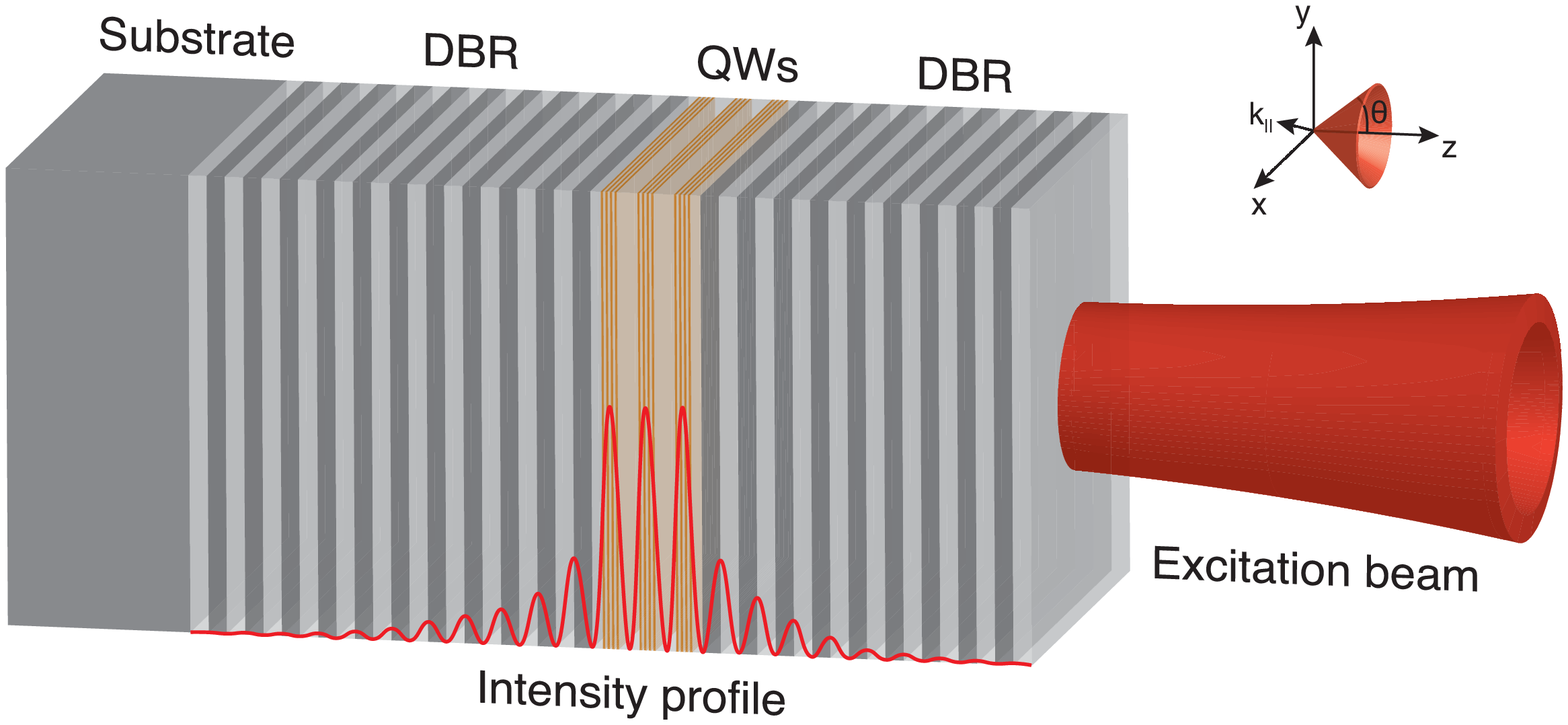}
   \caption{ (color online) 3D illustration of microcavity structure used in this work. The red lines indicate the intensity distribution of the confined optical field. The red cone shows the pump laser beam. The dark and light gray alternate layers are the distributed Bragg reflectors (DBRs) that are used to confine the light in the cavity, shown as the orange region. Quantum wells (QWs) are shown as the orange layers. }
\label{microcavity3d}
\end{figure}

\begin{figure}[htbp]
\centering
  \includegraphics[width=0.5\textwidth]{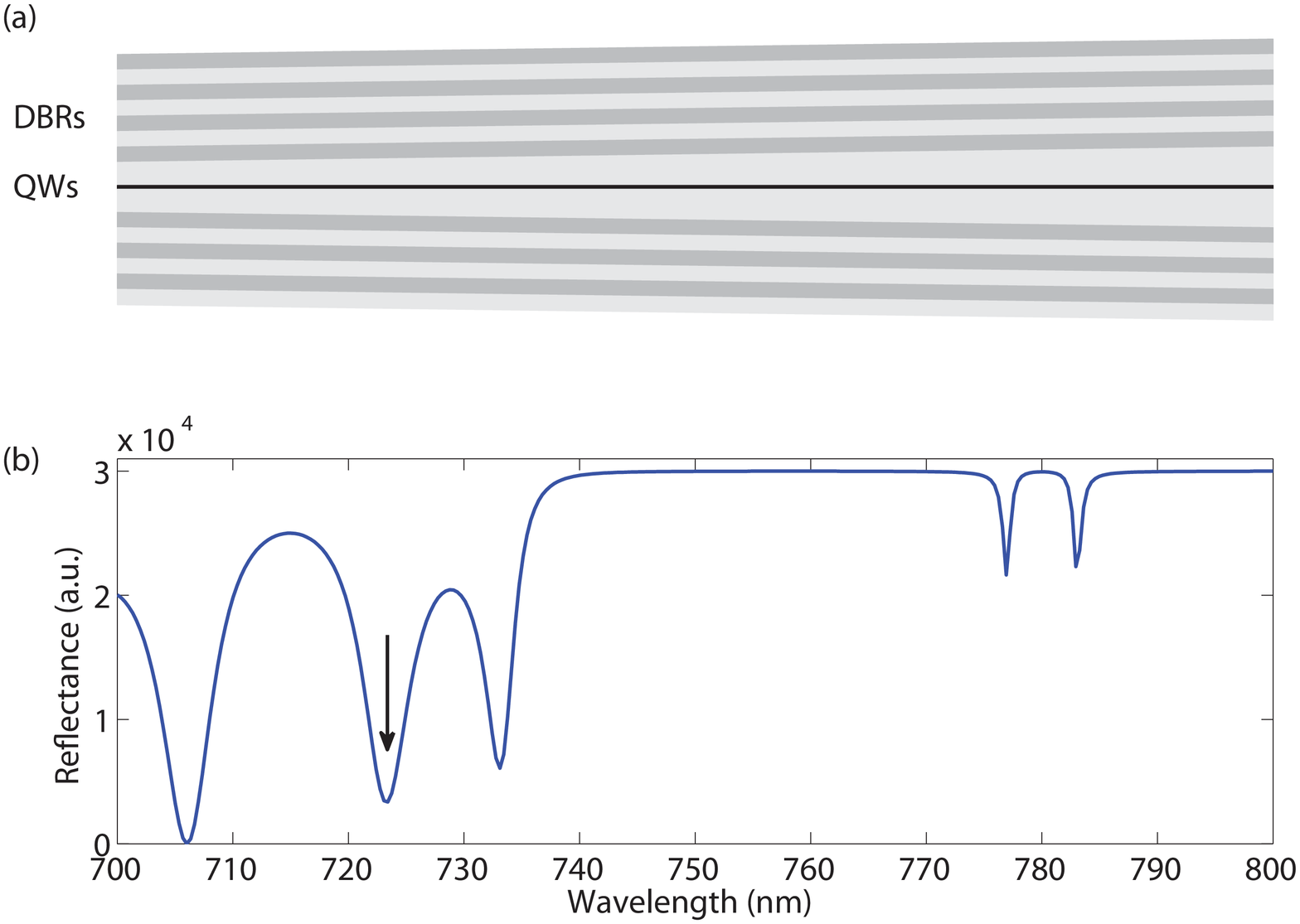}
   \caption{ (color online) Schematic illustration of microcavity structure used in this work. The dark and light gray alternative layers indicate the distributed Bragg reflectors (DBRs) that are used to confine the light in the cavity. Sets of quantum wells (QWs) are shown as the black lines. The wedged  cavity allows different cavity resonance frequencies to be selected. (b) Calculated white light reflectance of the sample structure. The black arrow indicates the second reflection minimum above stop band to which the laser wavelength was tuned to match.}
\label{microcavity}
\end{figure}

\noindent\textbf{\textrm{Detailed sample structure}}\hspace{10pt} The structure of the sample used in the experiments described in the main text is shown in Fig.~\ref{microcavity3d} and Fig.~\ref{microcavity}a: three sets of four 7-nm GaAs quantum wells (QWs) are embedded at the three antinodes in a $3\lambda/2$ microcavity, with the front and
 back distributed Bragg reflectors (DBRs), composed of 32 and 40 pairs, respectively, of Al$_{0.2}$Ga$_{0.8}$As/AlAs $\lambda/4$ layers. This microcavity structure is identical to semiconductor microcavity structures with short polariton lifetimes \cite{Balili2007} except that the number of layers in the DBRs is doubled. This leads to an significant increase in the lifetime of the trapped cavity field to about 135 ps, implying a polariton lifetime at $\delta = 0$ of 270 ps, which has been verified by transport measurements \cite{Steger2015}. 
 
In Fig.~\ref{microcavity3d},  the cavity region is indicated as the orange section, and red lines show a plot of the intensity distribution of the confined light modes in the cavity. We also show the ring-shaped optical excitation beam as the red cone on the right side of the structure. The microcavity structure is wedged along one direction as shown  in Fig.~\ref{microcavity}a, which allows tuning of photon resonance across the exciton resonance, thus allowing control of the exciton fraction in the polariton state. 
  
Although the concept of doubling the number of DBR layers is simple, the fabrication is not trivial, because it requires much longer fabrication times, approximately 30 hours of molecular-beam epitaxy (MBE), with tight control during the entire growth process.  If the growth process is not well controlled, inhomogeneities in the lower levels will be amplified in higher levels.

Fig.~\ref{microcavity}b plots the simulated spectrum of white light reflection of a microcavity structure. The stop band for optical transmission spans from 740 nm to 820 nm. Peaks around 780 nm arise from lower and upper polariton resonances. In the measurements, we tuned the laser wavelength to the second reflection minimum above the stop band, as indicated by the black arrow.  

\begin{figure}[htbp]
\centering
  \includegraphics[width=0.45\textwidth]{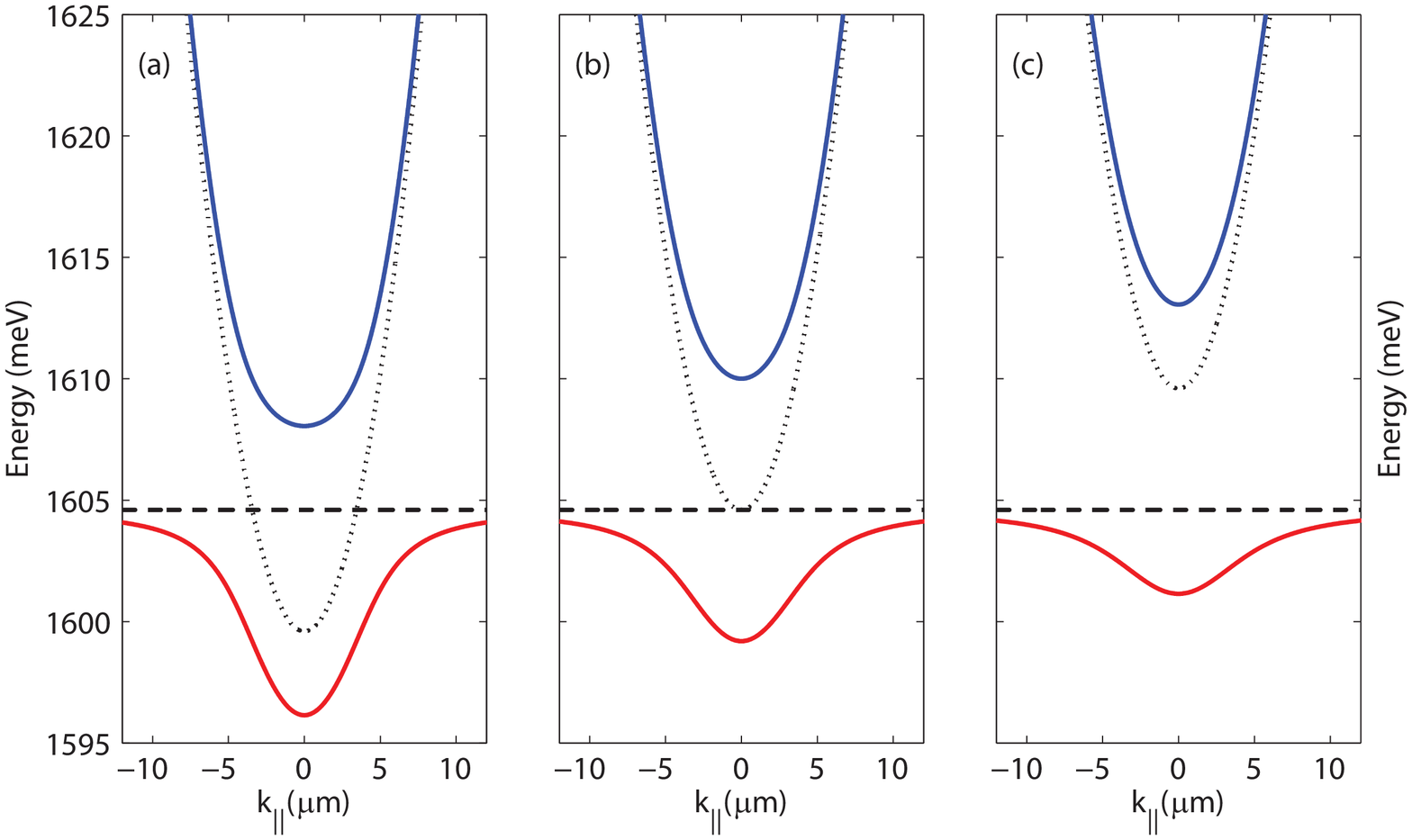}
   \caption{ (color online) Dispersion curves of polariton modes at three representative cavity detunings (a) $\delta = -5$ meV, (b) $\delta = 0$ meV, and (c) $\delta = 5$ meV. The dotted line shows the confined cavity mode, and the dashed line shows the bare exciton mode. The blue and red solid lines indicate the upper polariton (UP) and lower polariton (LP) branches, respectively, arising from the strong coupling between cavity modes and exciton modes. Our sample parameters were used in the calculations.}
\label{energy_dispersion}
\end{figure}

\vspace*{0.3cm}
\noindent\textbf{\textrm{Background on exciton-polaritons in semiconductor microcavities}}.\hspace{10pt}  Exciton-polaritons are formed in semiconductor microcavities through strong coupling between optical modes of the microcavity and exciton transitions of a material embedded inside the microcavity \cite{Kavokin2007, Deng2010, Snoke2010, Carusotto2013}. For the case of a single microcavity mode and a single exciton transition, two polariton modes, the upper and lower polaritons, are formed with energies $E_{LP}(k_{||})$ and $E_{UP}(k_{||})$ given by:
\begin{align}
E_{LP/UP}(k_{||}) = \frac{1}{2}\left[E_{X}({k_{||}}) + E_{C}(k_{||}) \mp \sqrt{\Omega^2 + \delta^2(k_{||})}\right]
\label{eq:Epolariton}
\end{align}
where $k_{||}$ is the wave vector in the plane perpendicular to the microcavity confinement direction, $E_X(k_{||})$ is the energy of the exciton transition, $E_C(k_{||})$ is the energy of the cavity mode, $\delta(k_{||})$ is the detuning energy defined as $\delta(k_{||})=E_C(k_{||})-E_X(k_{||})$, and $\Omega$ is the strength of radiative coupling between the exciton and cavity field, also known as full Rabi splitting energy. The confinement of light gives the cavity mode a parabolic dispersion in the plane perpendicular to the confinement direction: $E_C\simeq E_C(0) + \hbar^2k_{||}^2/2m_C$, where $m_C$ is the effective mass of the cavity field. This effective mass is typically $10^{-4}$ times lighter than the vacuum electron mass, and about $10^{-3}$ times less than an exciton in a GaAs quantum well structure, so that $E_X(k_{||})$ is essentially constant with $k_{||}$. The energies $E_X(k_{||}),E_C(k_{||}),E_{LP}(k_{||})$, and $E_{UP}(k_{||})$ are given in Fig.~\ref{energy_dispersion} for three different values of $\delta(k_{||}=0)$. The energies were calculated using (\ref{eq:Epolariton}) and parameters matching the sample structure used in the experiments: $\Omega = 10.84$ meV and $E_X(0)=1604.6$ meV. 

\begin{figure}[htbp]
\centering
  \includegraphics[width=0.40\textwidth]{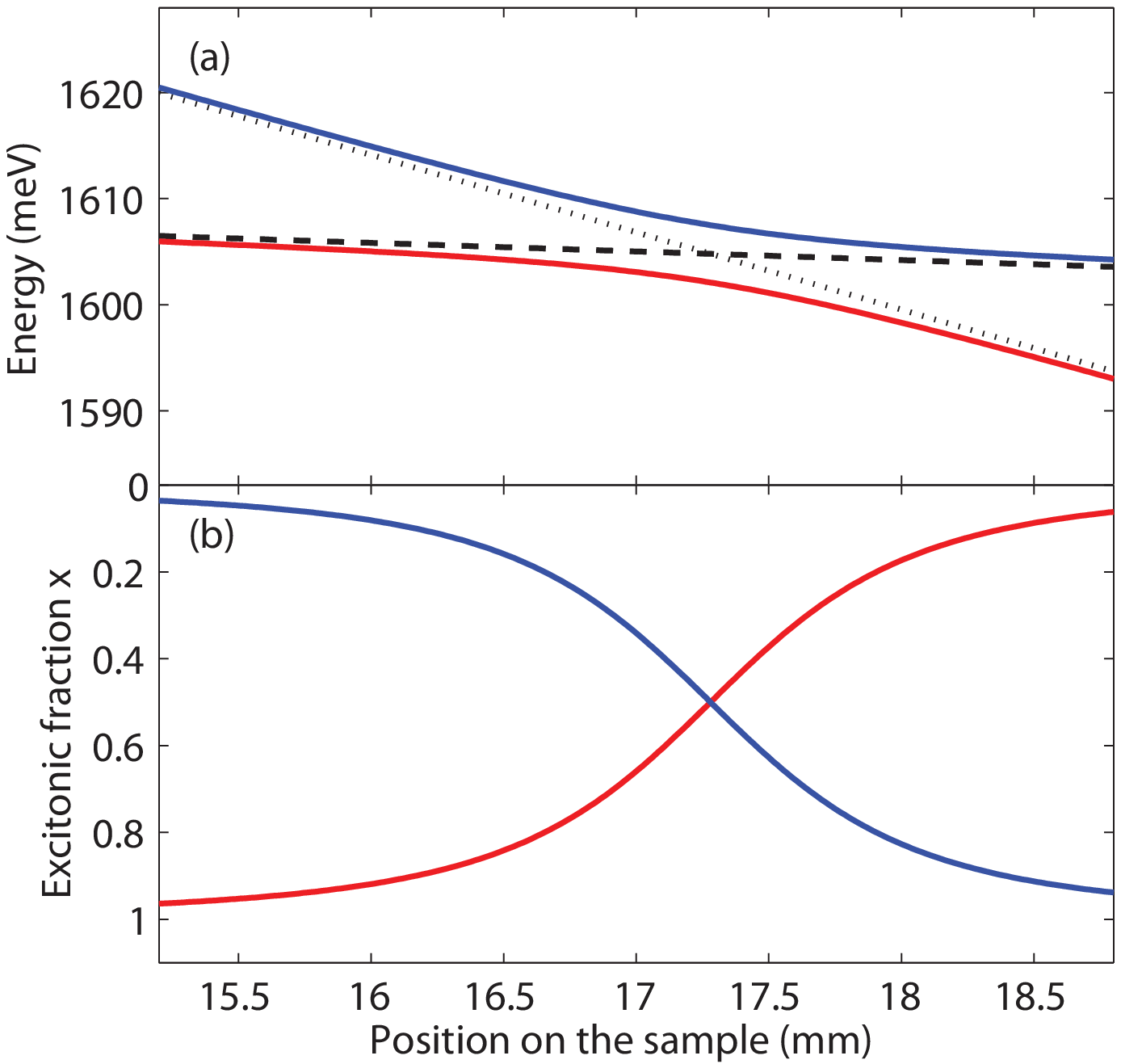}
   \caption{ (color online) (a) The calculated upper polariton (blue line) and lower polariton (red line) ground state ($k_{||}=0$) energies at different positions of the sample. The dashed line indicates the exciton energies, and the dotted line shows the cavity energies. (b) Excitonic fractions of upper polaritons (blue line) and  lower polaritons (red line) at different sample positions.}
\label{spatial_dispersion}
\end{figure}

The length of the cavity increases monotonically along one direction of the QW plane so that the energy of the cavity mode can be tuned relative to the exciton resonance energy, as shown in Fig.~\ref{spatial_dispersion}, allowing us to experimentally tune $\delta(k_{||}=0)$. The energies of all modes in Fig.~\ref{energy_dispersion} are plotted as a function of $k_{||}$, the in-plane wave vector. 

The polariton modes are linear superpositions of the exciton and microcavity photon modes. The lower polariton and upper polariton operators, $\hat{P}_{k_{||}}$ and $\hat{Q}_{k_{||}}$, respectively, can be written in terms of exciton and cavity operators, $\hat{a}_{k_{||}}$ and $\hat{b}_{k_{||}}$:
\begin{align}
\hat{P}_{k_{||}} &= X(k_{||})\hat{a}_{k_{||}}+C(k_{||})\hat{b}_{k_{||}k_{||}}\\
\hat{Q}_{k_{||}} &=-C(k_{||})\hat{a}_{k_{||}}+X(k_{||})\hat{b}_{k_{||}}.
\end{align}
The coefficients, $X(k_{||})$ and $C(k_{||})$, are called the exciton and cavity Hopfield coefficients \cite{Hopfield1958} and are given by 
\begin{align}
|X(k_{||})|^2 &= \frac{1}{2}\left(1+\frac{\delta(k_{||})}{\sqrt{\delta^2(k_{||})+\Omega^2}}\right)\\
|C(k_{||})|^2 &= \frac{1}{2}\left(1-\frac{\delta(k_{||})}{\sqrt{\delta^2(k_{||}k_{||})+\Omega^2}}\right).
\end{align}

The characteristics of the polariton modes are determined by the coefficients, which depend on $\delta(k_{||})$. The lower polariton is more photon-like and the upper polariton is more exciton-like for $\delta(k_{||})<0$, and the lower polariton is more exciton-like and the upper polariton is more photon-like when $\delta(k_{||})>0$. Due to the wedge in the cavity thickness, we can easily tune the excitonic fraction $|X(k_{||})|^2$ of lower polaritons by moving the excitation spot to different sample positions, as shown in Fig.~\ref{spatial_dispersion}b, where we plot $|X(k_{||}=0)|^2$ at different positions on the sample. As seen in Fig.~\ref{energy_dispersion} and Fig.~\ref{spatial_dispersion}, the energies and shapes of the polariton dispersion curves depend strongly on $\delta$:
positive detuning results in lower polaritons that are more exciton-like, with a heavier effective mass and stronger interactions with phonons and other carriers, while negative detuning results in lower polaritons that are more photon-like, with a smaller effective polariton mass and weaker interactions with phonons and other carriers.

\vspace*{.3cm}
\noindent\textbf{\textrm{Review of earlier thermalization data.}} In order to justify the realization of Bose-Einstein condensation,  macroscopic occupation of a single quantum state, which translates to the macroscopic quantum coherence, together with an equilibrium Bose-Einstein distribution of particles, which is crucial in deriving the temperature of the system and the phase diagram of the condensation, need to be verified \cite{Moskalenko2000, Pethick2002, Pitaevskii2003}. Various systems including a simple laser as well as short-lifetime microcavity polaritons have shown the spontaneous emergence of macroscopic quantum coherence, however, a lacking of equilibrium distribution across the condensation threshold disqualifies these systems from being named as Bose-Einstein condensation. In the past decade, the polariton condensates have been widely described as a nonequilibrium Bose-Einstein condensation. This leads to not only a simple nomenclature problem, but also a failure of applying to the polariton condensates the well accepted knowledge in the atomic condensates, not mentioned new difficulties in disentangling the effect between the many-body renormalization and nonequilibrium in polariton condensates. It is therefore crucial to investigate whether polaritons and their condensates can thermalize to form an equilibrium distribution.

Two earlier works have addressed in depth the thermalization of polaritons in short-lifetime samples. In Ref.~\cite{Kasprzak2008}, the polariton condensate in CdTe-based samples was studied over a wide range of lattice temperatures and polariton densities. In the reported data, the distributions $ N(E)$ were of two types. At high temperature at low density, the distributions fit a classical Maxwell-Boltzmann distribution. At low temperature and high density, the distributions  had a peak at $E=0$,  but could not be fit by an equilibrium Bose-Einstein distribution with a well-defined temperature and chemical potential. For example, in the curve for $T$ = 5.3 K in Fig.~1a of Ref.~\cite{Kasprzak2008}, the best fit to an equilibrium Bose-Einstein distribution {\it\bf missed the data by about a factor of 10} at low energy, leading to the difficulty in extracting the chemical potential, and also the temperature, at the onset of condensation. There was no comparison of the data at different densities, so that the meaningfulness of the fit values of chemical potential (in particular, the prediction of the absolute number of particles from the chemical potential) could not be verified. The interaction strength of the polaritons was varied by changing the detuning, and it was shown that the distributions became much further from equilibrium as the interaction strength decreased. The authors thus conclude that polariton condensation is driven by kinetics and cannot be termed as Bose-Einstein condensation (which is thermodynamic phase transition) given the nonequilibrium particle distributions.  

In Ref.~\cite{Deng2006}, the interaction strength was also varied by changing the detuning, and it was shown that for weak interactions, the distribution $N(E)$ was {\bf far from equilibrium}, while for strongly excitonic detuning (6 to 9 meV), the distribution became more thermalized. Reasonable fits to a Bose-Einstein equilibrium distribution were reported, but the system as a whole was still strongly in nonequilibrium. An intense pulsed laser was used, which gave a rapidly changing temperature and density of the polaritons on time scales of tens of picoseconds. It could be argued, however, that on short time scales of 1-2 picoseconds, the gas could be considered to be in quasi-equilibrium. More significantly, the strongly excitonic nature of the polaritons gave them a very short mean free path, which implies a very low diffusion constant. Since the polaritons were generated with a single laser spot, this led to an inhomogeneous spatial distribution. The momentum-space data were collected by integrating over the entire spatial profile. In this type of experiment with short-lifetime polaritons and a single Gaussian laser excitation spot, the potential-energy profile felt by the polaritons is strongly renormalized, leading to both self-trapping of the condensate in a small, quasi-harmonic potential inside the laser spot \cite{Deng2003}, and free streaming away from the spot \cite{Richard2005}. In general, it has been shown that integration of a spatially inhomogeneous distribution can lead to {\bf misleading fits to a Bose-Einstein distribution} \cite{OHara2000}, although reasonable assumptions for density and temperature variations can give fits with an average temperature and chemical potential \cite{Snoke1990}. Later work by the same group \cite{Roumpos2012} with the same short-lifetime structure used a large laser spot with a flat intensity profile and resonant detuning that gave longer mean free path to reduce the spatial inhomogeneity. In this configuration, the spatial coherence properties were measured. The power law for the spatial coherence was found to depend crucially on the nonequilibrium nature of the polariton gas \cite{Roumpos2012, Dagvadorj2015}.

These early attempts in addressing the thermalization behavior of polaritons and their condensates serve an important step in understanding polariton condensation, however, it clearly does not resolve the debate, as terms such as ``instrinsic nonequilibrium character'' and ``far from equilibrium'' have still been widely used in the literatures in the past 2 years \cite{Byrnes2014, Dominici2015, Sanvitto2016, Altman2015,Voronova2015}.

\begin{figure*}[htbp]
\centering
\subfloat{\includegraphics[width=0.32\textwidth]{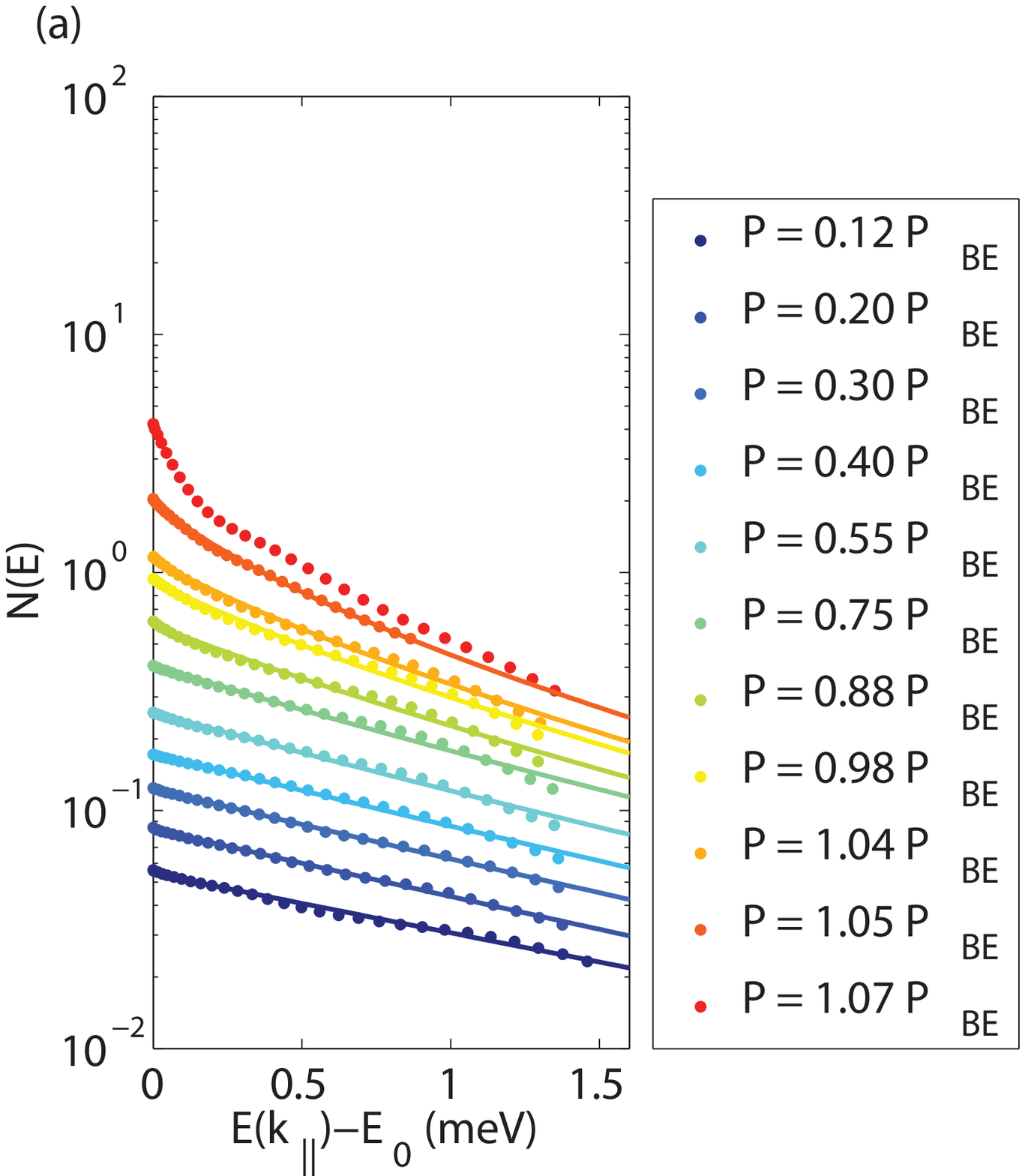}} 
\subfloat{\includegraphics[width=0.32\textwidth]{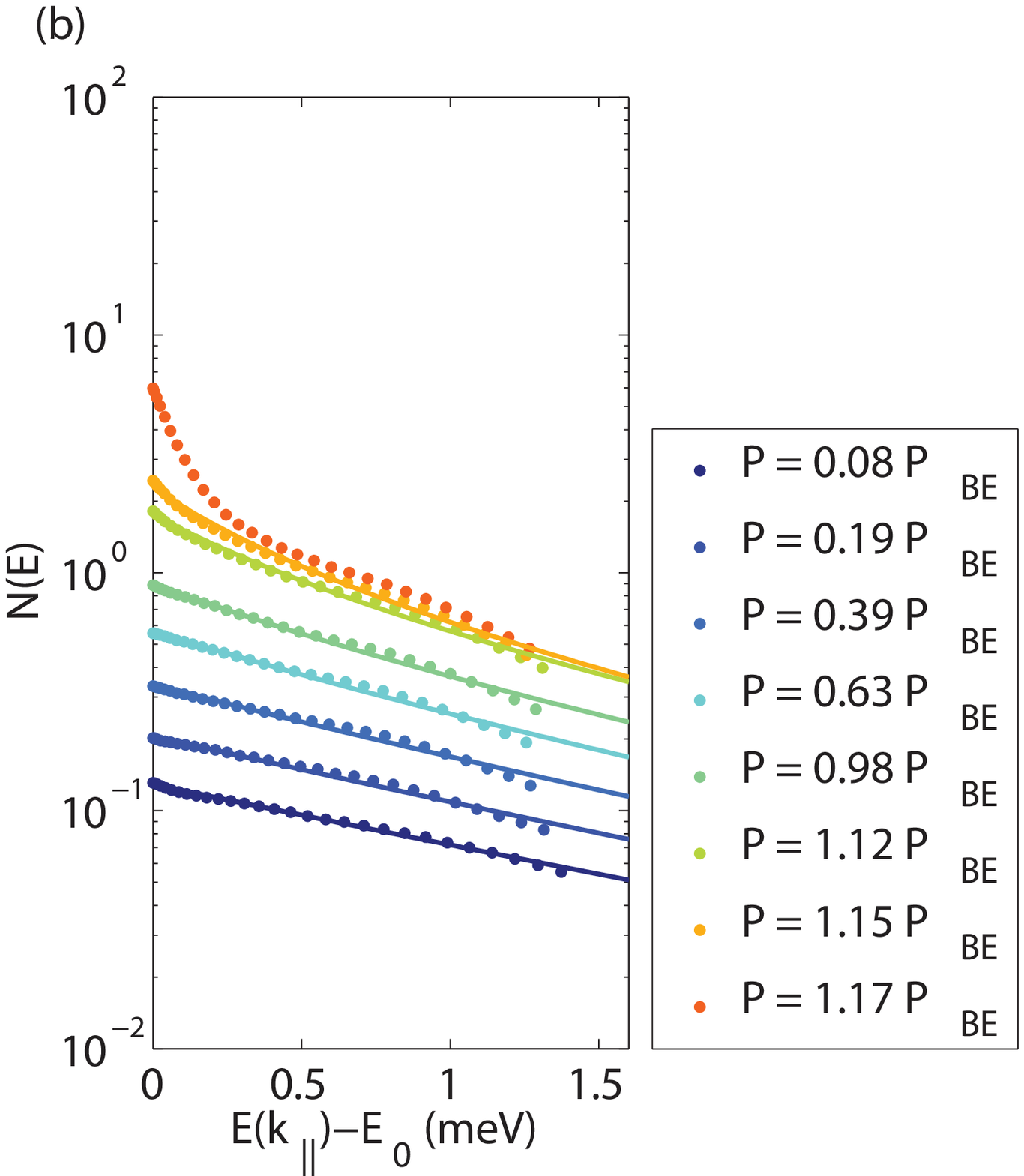}} 
\subfloat{\includegraphics[width=0.32\textwidth]{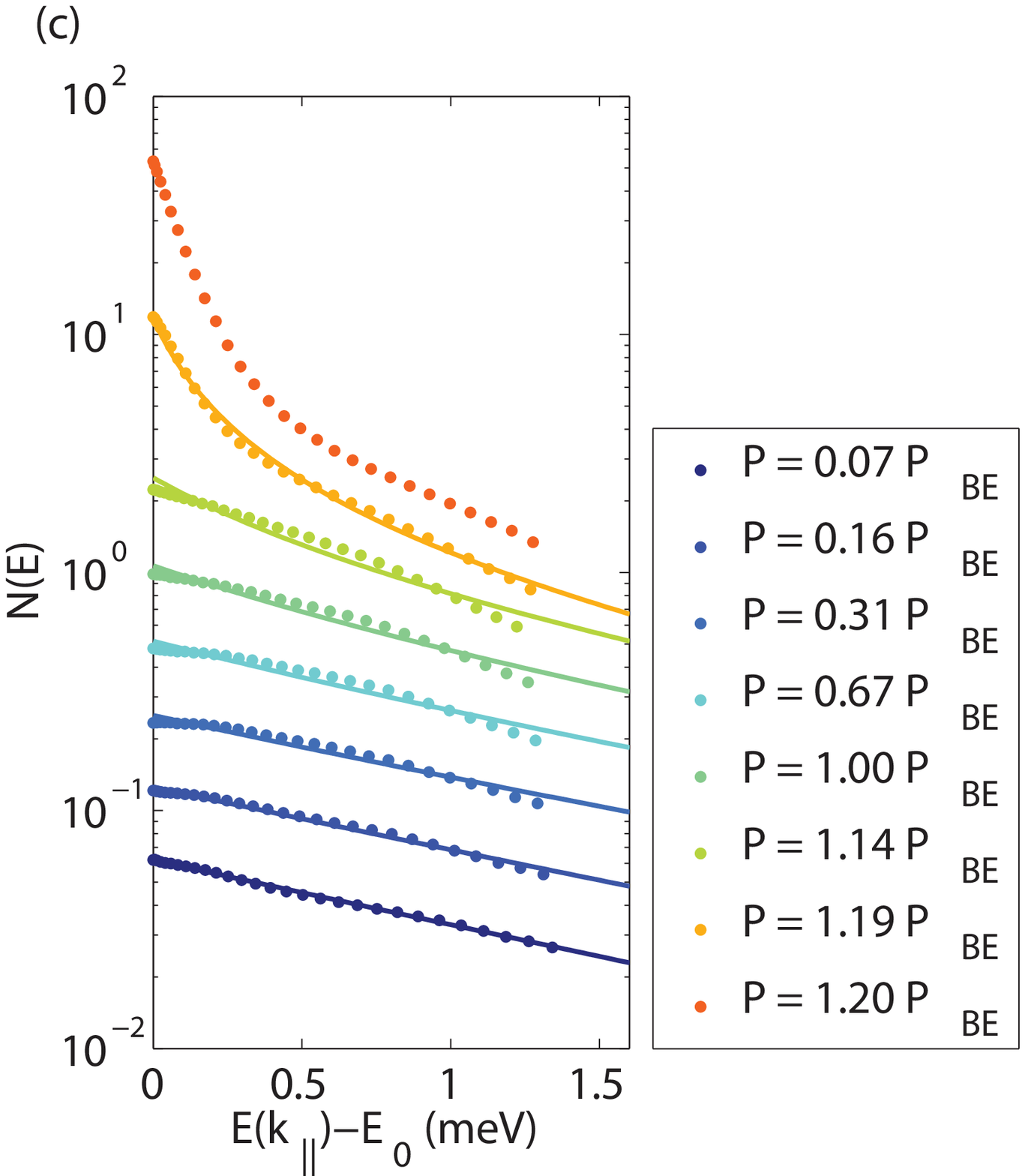}} 
\caption{Energy distributions of polaritons (dots) and the best fits to equilibrium Bose-Einstein model (solid lines)  at a bath temperature of $T=10.0$ K (a), $T=17.5$K (b) and $T=25.0$  K(c), with corresponding threshold pump powers $P_{BE}=435$ mW, $P_{BE} = 474$ mW, and $P_{BE} = 557$ mW, respectively. The cavity detuning is $\delta = 0$ meV, the same as that in Fig.~2b in the main text}
\label{Edistribution}
\end{figure*}

\vspace*{.3cm}
\noindent\textbf{\textrm{Energy distributions at different temperatures.}} In Fig.~\ref{Edistribution}, we show the energy distributions at three representative  bath temperatures, i.e., $T_{bath}=10.0$ K, $T_{bath}=17.5$ K and $T_{bath} = 25.0$ K, at the same cavity detuning as that of Fig.~2 in the main text, i.e., $\delta = 0$ meV, for a series of pump powers. As seen in these figures, the distributions at all temperatures fit well to the ideal Bose-Einstein function in Eq.~(1) in the main text, up to a ground-state occupation number $N(0) \sim$ 2--3. At higher densities, the distribution deviates from the ideal Bose-Einstein distribution. This is not surprising, since many-body effects play an important role when there is a large ground state occupation. The decrease in the energy range of the observed data as we increase the excitation power is a result of spectral narrowing, particularly when condensation forms, as shown in Fig.~2 in the main text.

\vspace*{.3cm}
\noindent\textbf{\textrm{Goodness of fit.}} Following Ref.~\cite{interactions}, we calculated polariton number in the field of view by integrating the fitted distribution 
\begin{align}
N_{fit} &= \frac{gmS}{2\pi\hbar^2}\int_0^{E_{max}}N(E)dE 
\end{align}
where $g = 2$ is the spin degeneracy factor of polaritons, $m$ is the effective mass of polaritons, $S$ is area of the field of view, and $E_{max}$ is the upper bound of the energy limit collected by the objective lens. We also computed the polariton number in the field of view by directly summing up the CCD count per second multiplied by the appropriate conversion factor, namely
\begin{align}
N_{int} &= \sum_i\frac{N_i\tau_i}{\xi M_i} \label{sum_CCD}
\end{align}
where $N_i,\tau_i$ and $M_i$ are the CCD counts per second, the lifetime of the polariton state, and the density of $k$ states at pixel $i$. $\xi$ accounts for the overall optical efficiency of the collection setup, including the loss in the dispersive grating and the quantum efficiency of the CCD. We define the goodness of fit as 
\begin{align}
gof &= \frac{N_{fit}^c}{N_{int}^c}
\end{align}
\begin{figure}[htbp]
\centering
  \includegraphics[width=0.30\textwidth]{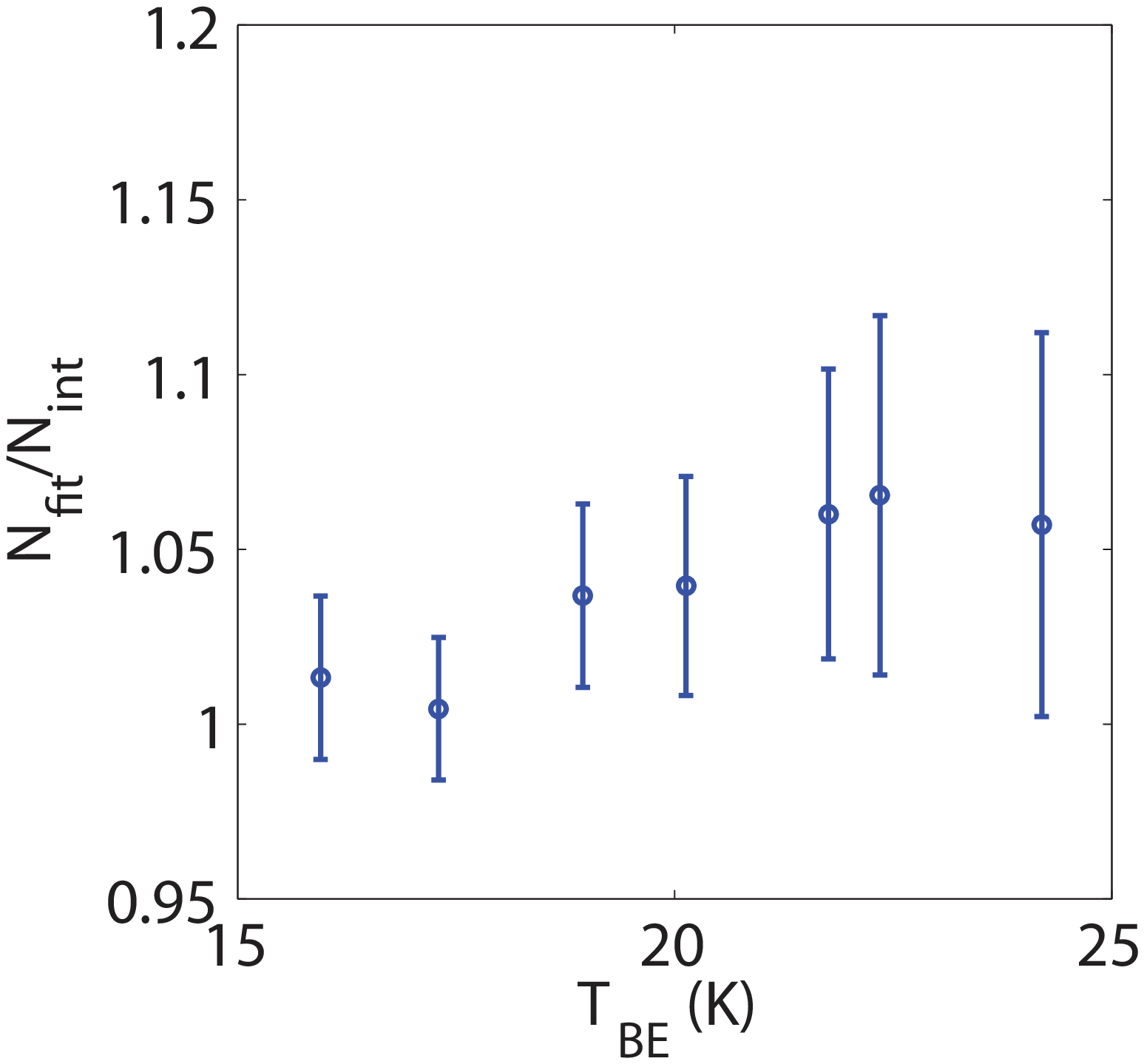}
   \caption{ (color online) Goodness of fit at different bath temperatures.}
\label{gof}
\end{figure}
where the superscript $c$ indicates the critical point of Bose degeneracy determined from main text. This quantity can be used as a quantitative measure of the overall quality of Bose-Einstein fitting. Fig.~\ref{gof} plots the $gof$ at the critical threshold $\mu/k_BT = -\ln 2$ for the data shown in Fig.~5b of the main text. As can be seen, this factor varies within $5\%$ for most of the temperatures. This confirms the high-quality fits, and thus confirms full thermalization of polariton gas in the range of studied temperatures. The slight increase in the $gof$ might come from the temperature-dependent emission rate of polaritons, which was not taken into account in the current data analysis routine. 

\vspace{.3cm}
\noindent\textbf{\textrm{Replot of phase diagram.}} Fig.~5b in the main text shows the phase diagram deduced by fitting the $N(E)$ data to a Bose-Einstein distribution, and then integrating the fit curve using Eq.~(2) in the main text. This method allows us to account for the high-energy tail of the particle distribution which extends outside of our detected range, due to the fact that high-energy states correspond to high $k_{\|}$, which give light emission outside the numerical aperture of our collection lens. Fig.~S6 shows two alternate ways of deducing the phase diagram, which rely less on the theoretical fit. 

\begin{figure}[htbp]
\centering
  \includegraphics[width=0.50\textwidth]{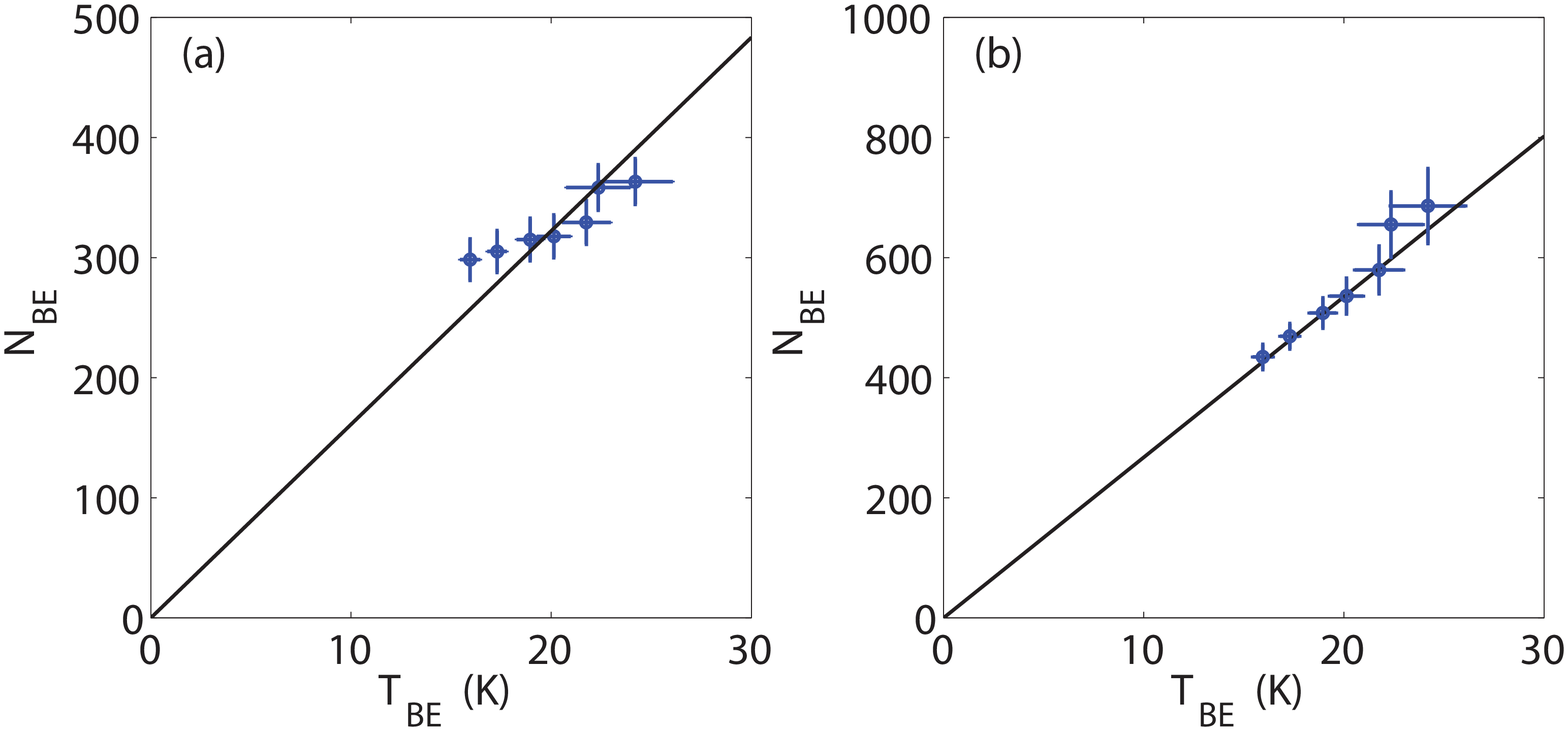}
   \caption{ (color online) $N_{BE}$ in the field of view versus $T_{BE}$, with $N_{BE}$ computed from only summing the CCD counts in the field of view (a) and including high-energy corrections (b). Solid lines are best linear fits $N_{BE}\propto T_{BE}$. }
\label{phase_diagram}
\end{figure}

Fig.~\ref{phase_diagram}(a) plots the total number of particles computed from summing the CCD counts per second as in Eq.~(\ref{sum_CCD}), without using any fit.
The black line is the best linear fit $y\propto x$. As can be seen, this plot deviates from linearity. This deviation is a result of the change in the fraction of polaritons in the field of view at different temperatures. As the temperature of the polaritons increases, the energy distribution tends to become broad, as indicated by a reduction in the slope in the semilog plot of $N(E)$ in Fig.~2 in the main text, and this will lead to more polaritons outside of the field of view of our measurements. 

Fig.~\ref{phase_diagram}b shows a correction to Fig.~\ref{phase_diagram}a, using the integrated CCD counts within the field of view, but adding a correction for the high-energy tail outside our field of view, based on the fits to the Bose-Einstein distribution.  The particle numbers plotted in Fig.~\ref{phase_diagram}b are therefore
\begin{align}
N_{tot} &= N_{int}+ \frac{gmS}{2\pi\hbar^2}\int_{E_{max}}^{E= \textrm{20 meV}}N(E)dE 
\end{align}
where $N_{int}$ is determined using Eq.~(\ref{sum_CCD}), $E_{max}$ is the maximum of observed energy range in the polariton distributions,  and the temperature and chemical potential in $N(E)$ are taken from the fitted values of Bose-Einstein distributions. The polariton population beyond the upper bound $E=20$ meV of the integration is negligible. The black line is a fitted linear relation. The difference is negligible compared to Fig.~5 in the main text, which is not surprising because of the fits to the Bose-Einstein distribution fall on top of the data at all temperatures. Compared to Fig.~\ref{phase_diagram}a, the proportionality is improved because of the inclusion of polariton populations at high energies which are outside of our field of view.

\end{document}